\documentclass[aps,prl,twocolumn,superscriptaddress,showpacs]{revtex4}
\usepackage{graphics}
\usepackage{bm}

\begin{document}

\title{Occupation of a resonant level coupled to a chiral Luttinger
  liquid}

\author{A. Furusaki} 

\affiliation{Yukawa Institute for Theoretical Physics, Kyoto University,
  Kyoto 606-8502, Japan}

\author{K.A. Matveev} 

\affiliation{Department of Physics, Duke University, Box 90305, Durham, NC
  27708}

\date{\today}

\begin{abstract}
  We consider a resonant level coupled to a chiral Luttinger liquid which
  can be realized, e.g., at a fractional quantum Hall edge.  We study the
  dependence of the occupation probability $n$ of the level on its energy
  $\varepsilon$ for various values of the Luttinger-liquid parameter $g$.
  At $g<1/2$ a weakly coupled level shows a sharp jump in $n(\varepsilon)$
  at the Fermi level.  As the coupling is increased, the magnitude of the
  jump decreases until $\sqrt{2g}$, and then the discontinuity in
  $n(\varepsilon)$ disappears.  We show that $n(\varepsilon)$ can be
  expressed in terms of the magnetization of a Kondo impurity as a
  function of magnetic field.
\end{abstract}


\maketitle

The transport of electrons in low-dimensional structures is strongly
affected by electron-electron interactions.  This effect is particularly
strong in one-dimensional (1D) systems, where the interacting electrons
form the so-called Tomonaga-Luttinger liquid \cite{Mahan}.  The main
feature of this system is the power-law suppression of the tunneling
density of states at low energies.  This effect has been recently observed
experimentally in 1D electron systems realized in quantum Hall edges
\cite{Chang}, carbon nanotubes \cite{Bockrath,Yao}, and quantum wires
\cite{Auslaender}.

A generic property of mesoscopic devices is the presence of imperfections
that affect the transport of electrons.  The most important are the
defects that form energy levels near the Fermi level and give rise to
strong scattering of electrons.  In this paper we consider a
\textit{single} level coupled to a 1D system exhibiting Luttinger
liquid behavior.
Experimentally the level can be formed either by a random impurity or by
an artificially created quantum dot in the vicinity of the 1D system.  In
recent experiments \cite{Grayson,Auslaender} resonant tunneling via such a
level has been observed.  The theory of resonant tunneling via a level
coupled to two 1D conductors was developed in
Refs.~\cite{Kane,Chamon,Furusaki}.

In contrast to the previous theoretical work \cite{Kane,Chamon,Furusaki},
we study a resonant level coupled to a single 1D conductor.  The main
quantity of interest is the occupation probability $n$ of the level as a
function of its energy $\varepsilon$ (measured from the Fermi level).
Clearly, at large positive energy $\varepsilon$ the level is empty, $n=0$,
and at large negative energy it is filled, $n=1$.  If the level is coupled
to a Fermi-liquid lead, the occupation probability changes continuously
from 0 to 1 when the energy $\varepsilon$ is within the level width
$\Gamma$ from the Fermi level:
\begin{equation}
  \label{eq:Fermi_liquid}
  n(\varepsilon) = \frac{1}{2} -
                   \frac{1}{\pi}\arctan\frac{\varepsilon}{\Gamma}. 
\end{equation}
We show below that the dependence $n(\varepsilon)$ is modified
significantly in the case of coupling to a Luttinger-liquid lead.  Most
importantly, if the interactions are strong enough, $n(\varepsilon)$ has a
discontinuity at $\varepsilon=0$.  

We model the level coupled to a 1D conductor by the  Hamiltonian
$H=H_0+H_t$, where 
\begin{eqnarray}
  \label{eq:H_0}
  H_0&=&\varepsilon a^\dagger a 
    + \frac{\hbar v}{4\pi}\int_{-\infty}^{\infty}
                            \left(\frac{d\varphi}{dx}\right)^2 dx,\\
  \label{eq:H_t}
  H_t&=& t(a^\dagger \psi + \psi^\dagger a), 
          \quad \psi=\sqrt{\frac{D}{2\pi\hbar v}}\,e^{i\varphi(0)/\sqrt{g}}.
\end{eqnarray}
The first term in Eq.~(\ref{eq:H_0}) describes the resonant level with
energy $\varepsilon$; the fermion operator $a^\dagger$ creates an electron
in this state.  The second term in Eq.~(\ref{eq:H_0}) describes the
Tomonaga-Luttinger liquid in the lead in terms of a chiral boson field
$\varphi(x)$ with commutation relations $[\varphi(x),
\varphi(y)]=i\pi\,\text{sgn}(x-y)$.  The chiral Luttinger liquid model is
the most natural description of fractional quantum Hall edge states
\cite{Wen}.  In the case of a non-chiral 1D system coupled to the level at
only a single point $x=0$ one can neglect the ``odd'' bosonic modes and
convert \cite{Matveev} the Hamiltonian to the chiral form (\ref{eq:H_0}).
The Hamiltonian $H_t$ describes the tunneling of the electron between the
level and the point $x=0$ in the 1D system.  We use the standard bosonized
expression for the annihilation operator $\psi$ of an electron in the
chiral Luttinger liquid in terms of the field $\varphi(x)$, the
Luttinger-liquid parameter $g$, the bandwidth $D$ and the Fermi velocity
$v$.  In this paper we only consider the case of spinless electrons.  Many
of our results, including the existence of the discontinuity in
$n(\varepsilon)$ at $g<1/2$, are not sensitive to this assumption.  Also,
in the case of tunneling into edge states of quantum Hall system with
filling factor $g<1$ all electrons are spin-polarized by the magnetic
field.  In other cases the level can be polarized by applying the magnetic
field.

We first calculate the occupation probability $n(\varepsilon)$ in the
second-order perturbation theory in $H_t$.  At $\varepsilon>0$ the result
has the natural form
\begin{equation}
  \label{eq:second-order}
  n(\varepsilon) = \frac{\Gamma}
                        {\pi\nu_1(0)}  
                   \int_0^\infty 
                   \frac{\nu_g(\omega)}{(\omega+\varepsilon)^2}
                   d\omega.
\end{equation}
Parameter $\Gamma=t^2/2\hbar v$ has the physical meaning of the level
width in the Fermi-liquid case $g=1$.  The tunneling density of states
$\nu_g(\omega)$ of the Tomonaga-Luttinger liquid has the well-known form
\begin{equation}
  \label{eq:nu}
  \nu_g(\omega)=\frac{1}{2\pi\hbar v\Gamma(g^{-1})}
                  \left(\frac{|\omega|}{D}\right)^{\frac{1}{g}-1}
                  e^{-|\omega|/D},
\end{equation}
where $\Gamma(x)$ is the Gamma function.  Due to the electron-hole
symmetry of the problem, the occupation probability at negative energies
can be found as $n(\varepsilon)=1-n(-\varepsilon)$.  At $g=1$ the
expression (\ref{eq:second-order}) coincides with expansion of the
Fermi-liquid result (\ref{eq:Fermi_liquid}) at $\varepsilon\gg\Gamma$.

At $g>1/2$ the integral in Eq.~(\ref{eq:second-order}) diverges at
$\varepsilon\to 0$.  This means that in order to find $n(\varepsilon)$
near the Fermi level one has to sum up the whole series of the
perturbation theory in $t$.  On the other hand, at $g<1/2$ the integral is
convergent at all $\varepsilon$, i.e., for a weakly coupled level the
second-order perturbation theory is sufficient to describe
$n(\varepsilon)$ at any energy.  For instance, at $\varepsilon\to+0$, we
find
\begin{equation}
  \label{eq:n(0)}
  n(+0) = \frac{g^2}{\pi(1-g)(1-2g)}\,\frac{\Gamma}{D}.
\end{equation}
At $\varepsilon\ll D$ the occupation probability approaches its limit
(\ref{eq:n(0)}) following a power-law dependence:
\begin{equation}
  \label{eq:approach}
   n(+0) - n(\varepsilon) = 
   \begin{cases}{\frac{\Gamma}{\pi D}
    |\Gamma(\frac{2g-1}{g})|
    \left(\frac{\varepsilon}{D}\right)^{\frac{1}{g}-2},
    &  $\frac13<g<\frac12$,\cr
    \frac{\Gamma}{\pi D}
    \frac{\varepsilon}{D}\ln\frac{D}{\varepsilon}, &  $g=\frac13$,\cr
    \frac{2g^3}{(1-g)(1-2g)(1-3g)}
    \frac{\Gamma}{\pi D}
    \frac{\varepsilon}{D}, &  $g<\frac13$.
   }
   \end{cases}
\end{equation}
Since $n(+0)\neq n(-0)\equiv 1-n(+0)$,  the system reaches two different
ground states at $\varepsilon\to\pm0$, and the occupation probability
$n(\varepsilon)$ experiences a finite discontinuity at the Fermi level:
$\Delta\equiv n(-0)-n(+0)$.  The divergence of the perturbation theory at
$g>\frac12$ indicates that in this regime the coupling of the level to the
lead is a relevant perturbation that should lift the degeneracy of the
ground state at $\varepsilon=0$, resulting in continuous $n(\varepsilon)$.
This can be shown formally using the mapping to the Kondo problem
discussed below.

As the coupling $\Gamma$ of the level to the 1D system grows, the
magnitude of the jump $\Delta =1-2n(+0)$ at $g<\frac12$ decreases.  We
will now show that at a sufficiently strong coupling the jump disappears.
Let us consider a semi-infinite quantum wire in the region $x>0$ with a
scatterer at point $x=L$.  The system is described by the
Tomonaga-Luttinger model with a scatterer:
\begin{equation}
  H=\frac{\hbar v}{2\pi}\int_0^\infty
      \left[\frac{1}{g}\left(\frac{d\theta}{dx}\right)^2
           + g\left(\frac{d\phi}{dx}\right)^2\right] dx
     - w \cos(2\pi N).
  \label{eq:non-chiral}
\end{equation}
Here $N=[k_FL+\theta(L)]/\pi$ is the number operator of electrons
in the region $0<x<L$.  The fields $\theta$ and $\phi$ satisfy the
commutation relation $[\theta(x),\phi(y)]=(i\pi/2)[1+\text{sgn}(x-y)]$
and the boundary condition $\theta(x=0)=0$.

If the scatterer is very strong, $w\to\infty$, it effectively cuts the
region $0<x<L$ from the wire.  This region becomes a quantum dot, coupled
to the rest of the wire by tunneling through the barrier created by the
scatterer.  At large $w$ the field $\theta$ at point $x=L$ is pinned at
one of the values $\theta(L)=\pi m - k_FL$, where integer $m$ is the
number of electrons in the dot.  The ground state energy of the dot is
then $E_m=(\hbar v/2\pi g L)(\pi m - k_FL)^2$.  For a given $L$ the number
of electrons in the dot that minimizes $E_m$ is the integer nearest to
$k_FL/\pi$.  One can easily check that at $k_FL/\pi=
M+\frac12-\frac{gL}{\pi\hbar v}\varepsilon$ with integer $M$ and positive
$\varepsilon$ the ground state of the dot has $m=M$ electrons, and the
energy of adding $(M+1)$-st electron is $E_{M+1}-E_M=\varepsilon$.  Thus
at large $w$ the Hamiltonian (\ref{eq:non-chiral}) is equivalent to the
model (\ref{eq:H_0}) and (\ref{eq:H_t}) of a resonant level coupled to a
1D system \footnote{ The Hamiltonian (\ref{eq:non-chiral}) describing
  non-chiral bosons can be converted by a transformation
  $\varphi(x)=\sqrt{g}\, \phi(|x|) + \text{sgn}(x) \theta(|x|)/\sqrt{g}$
  to a chiral form similar to Eq.~(\ref{eq:H_0}).}.

One can now increase the coupling by lowering $w$.  In the limit of very
strong coupling, $w\to 0$, the scattering term in the Hamiltonian
(\ref{eq:non-chiral}) can be accounted for in first-order perturbation
theory.  In an infinite 1D system the thermal average of the scattering
potential shows power-law behavior at low temperatures as
$(T/D)^{g}w\cos(2k_FL)$.  However, if the distance $L$ from the scatterer
to the end of the wire is finite, the renormalization of the scattering
potential is cut off at the low energy scale equal to the level spacing in
the dot, $\sim \hbar v/L$.  Although the renormalization of the scattering
potential in a semi-infinite wire may be large, it remains finite, and at
sufficiently small $w$ one can use the first-order result $\delta E_0\sim
(\hbar v/LD)^{g}w\cos(2k_FL)$.  Thus at strong coupling the energy is an
analytic function of $k_FL$ which shows no singularities at degeneracy
points $k_FL=\pi(M+\frac12)$ where $\varepsilon=0$.  The analyticity of
$E_0(\varepsilon)$ means that the ground state is non-degenerate, and the
occupation probability $n(\varepsilon)=\partial E_0/\partial\varepsilon$
(see Eqs.~(\ref{eq:H_0}) and (\ref{eq:H_t})) is continuous.


\begin{figure}[tb]
 \resizebox{.35\textwidth}{!}{\includegraphics{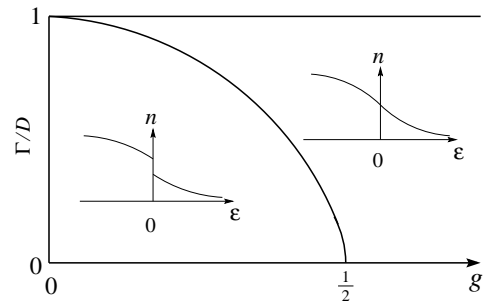}}
\caption{\label{figure}
  Schematic phase diagram for the behavior of the occupation probability
  $n(\varepsilon)$. The discontinuity occurs at \mbox{$g<\frac12$} and
  weak enough effective coupling $\Gamma/D$.}
\end{figure}

 From the perturbative calculations in the weak- and strong-coupling limits
we deduce a schematic phase diagram, Fig.~\ref{figure}.  The occupation
probability $n(\varepsilon)$ is a continuous function of $\varepsilon$ at
$g>\frac12$ (including the Fermi-liquid case $g=1$) and also at
$g<\frac12$, if the coupling to the lead is sufficiently strong.  On the
other hand, if $g<\frac12$ and the coupling is weak, $n(\varepsilon)$ has
a discontinuity at $\varepsilon=0$.  As the coupling is increased at
$g<\frac12$, the discontinuity $\Delta$ decreases until the phase boundary
is reached, above which $\Delta=0$.  We now show that $\Delta$ is not a
continuous function of the coupling strength, and that just below the
boundary $\Delta=\sqrt{2g}$.

Let us consider a level relatively weakly coupled to the lead with
$g<\frac12$.  The presence of a jump in the occupation probability
$n(\varepsilon)$ means that at $\varepsilon=0$ the ground state of the
system is degenerate.  One could expect that tunneling term (\ref{eq:H_t})
that couples the two degenerate states by moving an electron from the
level to the lead would lift the degeneracy.  This
does not always happen because the tunneling of an electron into a
Tomonaga-Luttinger liquid is suppressed at low energies, leading to
the renormalized level width
$\widetilde\Gamma(\varepsilon)\propto\varepsilon^\gamma$.  Thus at
$\gamma>1$ the coupling is irrelevant, as it is much smaller than the
level separation $\varepsilon$.  In the opposite case of $\gamma<1$ the
width is much larger than the level separation at $\varepsilon\to0$,
so the degenaracy of the ground state is lifted.  Thus at the phase
boundary in Fig.~\ref{figure} the exponent $\gamma=1$.

At $t\to0$ the suppression of tunneling manifests itself in the power-law
behavior of the density of states (\ref{eq:nu}).  Thus the
exponent $\gamma=\frac{1}{g}-1$, and the condition $\gamma>1$ of the
degenerate ground state reproduces our earlier result $g<\frac12$.
At $t\neq0$ the virtual hopping processes smear the quantization of the
charge, and at low energies the perturbation $H_T$ transfers only a
fraction $\Delta=n(-0)-n(+0)$ of an electron charge into the lead.
Mathematically this can be accounted for by a renormalized coupling term
$\widetilde H_T\propto t[a^\dagger
e^{i\varphi(0)\Delta/\sqrt{g}}+\text{h.c.}]$, c.f.\ Eq.~(\ref{eq:H_t}).
Thus the exponent of the tunneling density of states becomes
$\gamma= \frac{\Delta^2}{g}-1$ when only a fraction
$\Delta$ of the electron charge is transferred.  At the
phase boundary the condition $\gamma=1$ is then equivalent to
$\Delta=\sqrt{2g}$.

The phase diagram Fig.~\ref{figure} and the critical value of the jump
$\Delta=\sqrt{2g}$ at the phase boundary are the main results of this
paper.

One can also study how $n(\varepsilon)$ approaches the discontinuity at
$\varepsilon=0$ anywhere below the phase boundary in Fig.~\ref{figure}.
The discussion is essentially identical to the one leading to
Eq.~(\ref{eq:approach}) at $t\to 0$.  By substituting
$\nu(\omega)\propto\omega^\gamma$ instead of (\ref{eq:nu}) into
(\ref{eq:second-order}), we find at small positive $\varepsilon$,
\begin{equation}
  \label{eq:approach-2}
\frac{1-\Delta}{2}-n(\varepsilon)\propto
\varepsilon^{\min\{(\Delta^2/g)-2,1\}}
\quad
\text{for}~\Delta>\sqrt{2g}.
\end{equation}

Additional support for the above results is found by mapping our model to
two well-known problems: a two-level system with Ohmic dissipation and the
Kondo problem.  Let us define a unitary operator
$U_1=\exp[i\varphi(0)S^z/\sqrt{g}]$, where $S^z=a^\dagger a-1/2$.  Unitary
transformation of the Hamiltonian (\ref{eq:H_0}) and (\ref{eq:H_t}) yields
the Hamiltonian of the two-level system
\begin{equation}
  \label{eq:H(two-level)}
U^\dagger_1 H U_1=
H_0+t\sqrt{\frac{2D}{\pi\hbar v}}S^x
-\frac{\hbar v}{\sqrt{g}}S^z\frac{d\varphi(0)}{dx}
+\text{const},
\end{equation}
where $S^x=(a+a^\dagger)/2$.  The second term describes hopping of a
particle between the two levels, $S^z=\pm\frac12$, and the third term is
the coupling to harmonic oscillators with dissipation strength
$\alpha=1/2g$ \cite{Leggett}.  It is known that there is a quantum phase
transition between a phase where a particle is localized in one of the two
levels and another phase where it is delocalized.  In the limit $t\to0$
the transition occurs at $\alpha=1$ \cite{Leggett}, in agreement with our
weak-coupling result \cite{Buttiker}.  Our analysis suggests that as the
coupling $t$ is increased in the localized phase, the difference $\Delta$
of the averages $\langle S^z\rangle$ over the two ground states at
$\varepsilon=0$ will decrease from 1 to $\sqrt{2g}=\alpha^{-1/2}$, after
which the system will suddenly delocalize, $\Delta=0$.  To the best of our
knowledge, only a brief discussion \cite{Chakravarty} can be found in the
literature on the $\varepsilon$ dependence of the ground-state average
$\langle S^z\rangle$ at $\alpha=1-\eta$, and the result is consistent with
$\Delta=\alpha^{-1/2}$ to linear order in $\eta$.

To understand the connection to the Kondo problem, let us
introduce another operator $U_2=U_1\exp[-i\sqrt{2}\varphi(0)S^z]$.
Unitary transformation yields
\begin{eqnarray}
U_2^\dagger HU_2&=&
H_0
+t\sqrt{\frac{D}{2\pi\hbar v}}
 \left(S^+ e^{i\sqrt{2}\varphi(0)}
        +S^- e^{-i\sqrt{2}\varphi(0)}\right)
\nonumber\\
&&
-\hbar v\left(\frac{1}{\sqrt{g}}-\sqrt{2}\right)
  S^z\frac{d\varphi(0)}{dx}
+\text{const},
  \label{eq:H(Kondo-boson)}
\end{eqnarray}
where $S^+=a^\dagger$ and $S^-=a$.
Equation (\ref{eq:H(Kondo-boson)}) is also obtained by bosonizing
the anisotropic Kondo model describing electrons
scattered by a spin-$\frac{1}{2}$ impurity,
\begin{displaymath}
H_K=
\int\! \epsilon_k a_{k\mu}^\dagger a_{k\mu}^{}dk
+ \int\!\!\int\! J_i S^i \sigma^i_{\mu\nu}a_{k\mu}^\dagger a_{k'\nu}^{}
   dkdk'
+\varepsilon S^z,
 \label{eq:H(Kondo-fermion)}
\end{displaymath}
where $\epsilon_k=\hbar vk$, $\sigma^i$ is the Pauli matrix, and
summation over repeated 
indices is assumed ($\mu,\nu=\uparrow,\downarrow$ and $i=x,y,z$).
The Kondo couplings $J_i$ are given by
\begin{equation}
J_\perp=t\sqrt{\frac{2\pi\hbar v}{D}},\quad
J_z=2\pi\hbar v\left(1-\frac{1}{\sqrt{2g}}\right),
\label{eq:J}
\end{equation}
where $J_x=J_y=J_\perp$.  The occupation probability $n(\varepsilon)$ is
directly related to the magnetization $\mathcal{M}$ of the impurity spin
$\bm{S}$ coupled to the Zeeman field $\varepsilon$:
$n(\varepsilon)=\mathcal{M}(\varepsilon)+1/2$.  The phase diagram
Fig.~\ref{figure} is easily understood from the renormalization-group flow
diagram of the Kondo problem.  The phase of a smooth $n(\varepsilon)$
curve corresponds to the antiferromagnetic Kondo coupling,
$J_z>-|J_\perp|$.  In this case the impurity spin is completely
screened as the magnetic field $\varepsilon$ is turned off:
$\mathcal{M}(\pm 0)=0$.  At $g-\frac12\ll 1$ 
in the weak-coupling limit, $\Gamma\ll D$, one can easily obtain
$n(\varepsilon)$ from the Bethe ansatz results \cite{Bethe} for
$\mathcal{M}(\varepsilon)$.

The phase of a discontinuous $n(\varepsilon)$ curve in Fig.~\ref{figure}
corresponds to the ferromagnetic Kondo case, $J_z<-|J_\perp|$, where the
spin flip coupling $J_\perp$ is renormalized to zero.  One might then
naively expect full moment to appear:
$\mathcal{M}=-\text{sgn}(\varepsilon)/2$.  This is not correct, however,
as we have seen in Eqs.~(\ref{eq:n(0)}) and (\ref{eq:approach}).  In the
weak-coupling regime we may use one-loop scaling equations for
$j_i=J_i/(2\pi\hbar v)$,
\begin{equation}
\frac{dj_\perp}{dl}=2j_\perp j_z,\quad
\frac{dj_z}{dl}=2j_\perp^2,\quad
\frac{d\mathcal{M}}{dl}=-2j_\perp^2\mathcal{M},
  \label{eq:scaling}
\end{equation}
where $dl=-d\ln D$.
For $|j_\perp|\le-j_z$ we integrate them to
$l=\ln(D/\varepsilon)$ with initial condition $\mathcal{M}=-1/2$.
We find $2\mathcal{M}(\varepsilon)=-\exp(j_z-j_z^*)$, where
\begin{equation}
\frac{j_z^*}{c}=
\frac{j_z-c+(c+j_z)(\varepsilon/D)^{4c}}
     {c-j_z+(c+j_z)(\varepsilon/D)^{4c}}, \quad
c=\sqrt{j_z^2-j_\perp^2},
 \label{eq:J_z}
\end{equation}
in agreement with Ref.~\cite{Anderson}.  To lowest order in
$\frac{1}{2}-g$, this one-loop result is consistent with
Eq.~(\ref{eq:approach-2}).  In particular, at the transition point
$j_\perp=j_z$, the magnitude of the magnetization jump
$\Delta=\mathcal{M}(-0)-\mathcal{M}(+0)=1+j_z$ agrees with our conjecture
$\sqrt{2g}=(1-j_z)^{-1}$ in the first order in $j_z$.  Approach to the
zero-field value is logarithmically slow at the transition point:
$\mathcal{M}(\varepsilon)-\mathcal{M}(+0)\propto 1/\ln\varepsilon$ for
$\varepsilon>0$.  It is worth mentioning that in the ferromagnetic case
the magnetization is a nonuniversal quantity.  In particular, models with
different bandwidth cutoffs have values of $\mathcal{M}$ different by
$\sim j_z^2$.  To be precise, our mapping is to the bosonized Kondo model
(\ref{eq:H(Kondo-boson)}).

The magnetization jump in the ferromagnetic Kondo model is also related to
the problem of a classical Ising chain with inverse-square interaction
\cite{Anderson}, whose reduced Hamiltonian is $\beta H_I=-K\sum_{i>j}s_i
s_j/|i-j|^2$, $s_i=\pm1$.  The connection is established by expanding the
partition function of the Kondo model in powers of $J_\perp$, which is
then identified with that of the Ising chain at $K=1/4g$ with additional
short-ranged irrelevant interaction that depends on $J_\perp$ and
bandwidth cutoff.  The 1D Ising model with $1/r^2$ interaction has a
finite-temperature phase 
transition \cite{Anderson}.  Below the transition temperature, or when
$K>K_c$, the Ising spins are ordered with a nonvanishing order parameter
$\Psi\equiv K\langle s_i\rangle^2$.  The order parameter vanishes
discontinuously as the coupling $K$ is reduced through the $K_c$, a 1D
analog of the universal jump at the Kosterlitz-Thouless transition.  It
has been shown rigorously that $\Psi\ge1/2$ in the ordered phase
\cite{Aizenman}.  Although unproved rigorously yet, it is believed that
$\Psi=1/2$ at $K=K_c$, which has been confirmed recently by a large-scale
numerical simulation \cite{Luijten}.  This means that
$\mathcal{M}(-0)-\mathcal{M}(+0)=\langle s_i\rangle=\sqrt{2g}$,
in agreement with our analysis.
Note that the external parameter driving the phase transition is $K$ in
the Ising model whereas it is $\Gamma/D$ in our model of a level coupled
to a 1D lead.

So far we have ignored Coulomb repulsion between the resonant level and
the 1D lead.  Assuming that the interaction is short-ranged due to
screening effects, we may model it by $H_u=u\,S^z \frac{d\varphi(0)}{dx}$.
Since the unitary transformation gives $U_2^\dagger H_u U_2=H_u +
\text{const}$, the effect of the Coulomb repulsion is to change the Kondo
coupling, $J_z\to J_z+2\pi u$.  This gives rise to a shift of the phase
boundary in Fig.~\ref{figure} towards smaller $g$.  In particular, at
$t\to0$ the phase boundary is at $g=\frac{1}{2}(1+\frac{u}{\hbar
  v})^{-2}$.  Physically, this shift is due to Mahan's excitonic effect
\cite{Mahan} that enhances tunneling probability.

It was recently demonstrated \cite{Berman,Duncan} that the charge of a
quantum dot coupled to a Fermi-liquid lead can be measured with the aid of
an electrometer based on a single electron transistor.  To test our
theoretical predictions one could perform a similar experiment with a lead
in the Fractional Quantum Hall regime.  A small dot coupled to the edge
state in the lead would then play the role of the resonant level.  In this
setup one can vary the parameter $g$ of the chiral Luttinger liquid in the
edge state by adjusting the magnetic field.  The strength of coupling of
the dot to the edge state in a GaAs heterostructure can be tuned by
changing the appropriate gate voltage.

In conclusion, we have shown that in the region of parameters shown in the
phase diagram Fig.~\ref{figure} the occupation probability
$n(\varepsilon)$ of a resonant level of energy $\varepsilon$ coupled to a
Luttinger-liquid lead is a discontinuous function of $\varepsilon$.
The height of the discontinuity in $n(\varepsilon)$ at $\varepsilon=0$
reaches $\sqrt{2g}$ at the phase boundary.  This picture is supported
by mapping to a Kondo model, and interesting connections can also be
made to the problems  of the dissipative two-level system and
the classical Ising chain with inverse-square interaction.

\begin{acknowledgments}
  We are grateful to S.~Katsumoto for valuable discussions and to the
  Aspen Center for Physics where part of the work was performed.  The work
  of AF was supported by the Grant-in-Aid for Scientific Research
  No.~11740199 from the Japan Society for the Promotion of Science (JSPS).
  KAM acknowledges the support of the Sloan Foundation, the JSPS, and NSF
  Grant DMR-9974435.
\end{acknowledgments}


\begin{thebibliography}{99}
  
\bibitem{Mahan} For a review see, e.g., G. D. Mahan, \textit{Many-Particle
    Physics,} 3rd ed. (Kluwer, New York, 2000).
  
\bibitem{Chang} A.M. Chang, L.N. Pfeiffer, and K.W. West, Phys. Rev. Lett.
  {\bf 77}, 2538 (1996).

\bibitem{Bockrath} M. Bokrath {\it et al.}, Nature (London) {\bf 397},
  598 (1999).

\bibitem{Yao} Z. Yao {\it et al.}, Nature (London) {\bf 402}, 273 (1999).

\bibitem{Auslaender} O.M. Auslaender {\it et al.}, Phys. Rev. Lett. {\bf
    84}, 1764 (1999).
  
\bibitem{Grayson} M. Grayson {\it et al.}, Phys. Rev. Lett. {\bf 86}, 2645
  (2001).
  
\bibitem{Kane} C.L. Kane and M.P.A. Fisher, Phys. Rev. B {\bf 46}, 15233
  (1992).

\bibitem{Chamon} C. de C. Chamon and X.G. Wen, Phys. Rev. Lett. {\bf 70},
  2605 (1993).
  
\bibitem{Furusaki} A. Furusaki and N. Nagaosa, Phys. Rev. B {\bf 47}, 3827
  (1993); A. Furusaki, Phys. Rev. B {\bf 57}, 7141 (1998).

\bibitem{Wen} X.-G. Wen, Int. J. Mod. Phys. B {\bf 6}, 1711 (1992).
  
\bibitem{Matveev} See, e.g., K.A. Matveev, Phys. Rev. B {\bf 51}, 1743
  (1995).
  
\bibitem{Leggett} A.J. Leggett {\it et al.,} Rev. Mod. Phys. {\bf 59}, 1
  (1987).

\bibitem{Buttiker} The second-order result (\ref{eq:n(0)}) is also
  found in P. Cedraschi and M. B\"uttiker,
  Ann. Phys. (N.Y.) {\bf 289}, 1 (2001).

\bibitem{Chakravarty} S. Chakravarty, Phys. Rev. Lett. {\bf 49}, 681
  (1982).

\bibitem{Bethe} A.M. Tsvelick and P.B. Wiegmann, Adv. Phys. {\bf 32},
  453 (1983); N. Andrei, K. Furuya, and J.H. Lowenstein,
  Rev. Mod. Phys. {\bf 55}, 331 (1983).

\bibitem{Anderson} P.W. Anderson and G. Yuval, J. Phys. C {\bf 4}, 607 
  (1971).

\bibitem{Aizenman} M. Aizenman, J.T. Chayes, L. Chayes, and C.M. Newman,
  J. Stat. Phys. {\bf 50}, 1 (1988); J.Z. Imbrie and
  C.M. Newman, Commun. Math. Phys. {\bf 118}, 303 (1988).
  
\bibitem{Luijten} E. Luijten and H. Me{\ss}ingfeld, Phys. Rev. Lett. {\bf
    86}, 5305 (2001).

\bibitem{Berman} D. Berman, N.B. Zhitenev, R.C. Ashoori, and M. Shayegan,
  Phys. Rev. Lett. {\bf 82}, 161 (1999).

\bibitem{Duncan} D.S. Duncan {\it et al.}, Appl. Phys. Lett. {\bf 74},
  1045 (1999).

\end{thebibliography}
\end{document}